# Eliciting Preferences of Ridehailing Users and Drivers: Evidence from the United States

Prateek Bansal[1], Akanksha Sinha[1], Rubal Dua*[2], Ricardo Daziano[1]

[1]Department of Civil and Environmental Engineering, Cornell University, United States
[2]King Abdullah Petroleum Studies and Research Center (KAPSARC), P.O. Box 88550, Riyadh 11672, Saudi Arabia

*Corresponding author: rubal.dua@kapsarc.org

## Abstract

Transportation Network Companies (TNCs) are changing the transportation ecosystem, but micro-decisions of drivers and users need to be better understood to assess the system-level impacts of TNCs. In this regard, we contribute to the literature by estimating a) individuals' preferences of being a rider, a driver, or a non-user of TNC services; b) preferences of ridehailing users for ridepooling; c) TNC drivers' choice to switch to vehicles with better fuel economy, and also d) the drivers' decision to buy, rent or lease new vehicles with driving for TNCs being a major consideration. We use a unique sample (N= 11,902) of the U.S. population residing in TNC-served areas. Elicitation of drivers' preferences using a large sample is the key feature of this study. The population-weighted statistical analysis indicates that ridehailing services are mainly attracting personal vehicle users as riders, without substantially affecting demand for transit. Moreover, around 10% of ridehailing users reported postponing the purchase of a new car due to the availability of TNC services. Using a multinomial logistic regression, we find that that the likelihood of being a TNC user increases with the increase in age for someone younger than 44 years, but the pattern is reversed post 44 years. This change in direction of the marginal effect of age is insightful as the previous studies have reported a linear negative association. Moreover, older ridehailing users with higher household vehicle ownership who live in suburban areas are less likely to pool rides. On the supply side, 65% of TNC drivers who work daily indicated that driving for TNCs was a consideration in vehicle purchase decisions. We also find that households with postgraduate drivers who drive daily and live in metropolitan regions are more likely to switch to fuel-efficient vehicles. These findings can inform transportation planners and TNCs in developing policies to encourage ridepooling and to improve the average fuel economy of the TNC fleet.

# 1. Introduction

Ridehailing services such as Uber and Lyft (also known as Transportation Network Companies or TNCs) are rapidly changing the landscape of transportation. These companies have transformed the way people travel around cities. Schaller (2018) reveals that 2.61 billion passengers used TNCs in 2017 in the U.S., a 37% increase from 2016. This increase in the adoption of TNC services can be attributed to the ease of access using a smartphone application along with a higher availability compared to the regulated, traditional taxi services.

Proponents of TNC services emphasize positive impacts such as making travel easier by providing more travel options, especially in high-demand areas with uncertain and inefficient transit services (Alemi et al., 2018). There is also a possibility of partnerships between transit agencies and TNCs to improve the first and the last mile connectivity. In one such pilot project, Uber rides to and from a commuter rail station in a suburb of Orlando, Florida were subsidized in March 2016 (Shaheen and Chan, 2016). TNCs also provide users with an opportunity to give up vehicle ownership or reduce vehicle use. A few recent studies have provided evidence in support of this transition. Rayle et al. (2014) report that almost 40% of TNC users reduced driving due to the adoption of ridehailing services in San Francisco. Moreover, Clewlow and Mishra (2017) conclude that more frequent TNC users are more likely to reduce their household vehicle ownership. Similar to carsharing services, TNCs with ridepooling have immense potential to reduce vehicle miles traveled (VMT) and thus greenhouse gas emissions and congestion arising from personal auto use (Martin and Shaheen, 2011). With these benefits in mind and to obtain a wider appeal among price-sensitive travelers, Uber and Lyft introduced their pooling services (UberPool and LyftLine) in 2014, which were heavily subsidized. Recently, Uber also introduced a new pooling service, UberPool Express, at an approximately 50% lower price than Uberpool and 75% lower than UberX (Hawkins, A., 2015). Additionally, given the higher mileage of vehicles used for taxis and ridehailing, high fuel economy and alternative fuel vehicles hold promise for adoption by ridehailing drivers and associated fleet providers. These strategies can also contribute toward lowering energy use and environmental impact associated with passenger auto travel.

Contrary to ridehailing's potential for improving societal welfare, a few studies have shown an adverse effect of these services, including induced demand for travel and a reduction in transit ridership in certain areas. For instance, in a survey across seven major cities in the United States, Clewlow and Mishra (2017) find that 49% to 61% of ridehailing trips would not have been made at all or could have been made by walking, biking, or public transit.

In sum, transportation planners and policymakers are uncertain about the impact of TNC services on energy use, the environment, and traffic congestion (Conway et al 2018). Whereas a decline in VMT and vehicle ownership can reduce greenhouse gas emissions (as established in the carsharing literature[1]), induced demand can compensate for such gains. These system-level impacts are manifestations of individual-level decisions, namely choice of ridehailing over transit or drive alone, preference to entirely depend on TNC services rather than owning vehicles, and TNC drivers' inclination toward buying fuel-efficient vehicles, among many others. However, in the absence of rich data sources, metropolitan planning organizations are unable to quantify the effect of TNCs on individual-level travel decisions, and system-level questions thus remain unanswered.

---

[1] Greenblatt and Shaheen (2015) estimate an average reduction of 34% in greenhouse gas emissions per household because 25% to 71% of users avoided vehicle purchase due to the availability of carsharing services.

Previous studies have focused on understanding travelers' preferences to use or not use ridehailing services. These studies have shown that early adopters of TNC services are primarily well-educated young individuals who come from affluent families (Rayle et al., 2014; Clewlow and Mishra, 2017; Alemi et al., 2018). In the context of TNC drivers, some studies have touched upon driver safety (Feeney, 2015), driver wages (Berger and Frey, 2017), and sociodemographic characteristics (Kooti, 2017; Hall and Krueger, 2018), as well as personal attitude of individuals who are willing to become TNC drivers (Berliner and Tal, 2018). But, to the best of our knowledge, no previous study has explored preferences of TNC drivers (such as buying a vehicle with improved fuel economy), perhaps due to a lack of driver-level datasets.

This study takes an important step toward bridging this gap by analyzing a survey data, procured from Strategic Vision Incorporated, consisting of a sample (N= 11,902) of the U.S. population residing in TNC served areas. The survey included information on sociodemographic characteristics, personal attitudes toward adopting TNC services as a user or driver, and changes in travel mode and vehicle ownership preferences after using these services. Using this revealed preference data, we investigate association between sociodemographic characteristics and the following: a) individuals' preferences for being a rider, a driver, or a non-user of TNC services; b) preferences of ridehailing users for ridepooling; c) ridehailing drivers' choice to switch to vehicles with better fuel economy, and also d) TNC drivers' decision to buy, rent or lease a new vehicle with driving for TNCs being a major contributing factor. We fit multinomial logistic regressions to answer the first and binary logistic regressions to answer the remaining research questions. We also provide insights about a) variation in individuals' frequency to use ridehailing services across different trip purposes and in absence of their preferred travel mode; b) impact of using ridehailing on vehicle ownership and other mobility decisions; c) preference of public transit users to adopt TNC services for first/last mile connectivity; d) activities undertaken by TNC drivers during downtime; e) and average pick-up time, across surveyed TNC drivers.

The remaining of the paper is organized as follows: section 2 provides a review of the relevant literature; section 3 discusses the survey data and key insights from the descriptive statistics of the sample and succinctly describes the methodology used in this study; section 4 summarizes main results; and conclusions and future research are discussed in section 5.

## 2. Literature Review

This section summarizes the contextual literature on individuals' preferences to use TNC services and the subsequent impacts on their mobility decisions. We first discuss the evolution of TNCs and then describe the sociodemographic and geographic characteristics of individuals who have a higher tendency to use TNC services. We then review the literature on how these services are changing the landscape of urban travel patterns by affecting vehicle ownership preferences and demand for other travel modes. We conclude with a review of studies focusing on ridehailing drivers, followed by highlighting the research gap that we are addressing in this study.

In recent years, transportation has gone through a rapid transformation due to rapid deployment of emerging technologies such as smartphone and internet (Taylor et al., 2015). These technological advancements have mitigated geographic constraints and are the main drivers of TNCs rapid growth. As of 2016, ridehailing services are active in almost 500 cities in the United States (Murphy, 2016). However, these services are still not a regular mode of transport. From a survey of 4,787 American adults, the Pew Research Center finds that only 3% and 12% of TNC riders use these services on a daily or a weekly basis, respectively (Smith, 2016).

Some correlation studies have identified the characteristics of travelers with a higher propensity to use ridehailing services. Results of a survey across seven major US cities indicate that the adoption rate of these services is almost double among college-educated individuals as compared to those without a college degree (Clewlow and Mishra, 2017). Furthermore, travelers younger than 29 years and older than 65 years are found to be the most and least frequent users of ridehailing services, respectively. Kooti et al. (2017) also obtain similar findings by analyzing Uber data that was collected over a span of seven months and contains 59 million rides and 4.1 million riders. The authors conclude that whereas younger riders are more likely to take frequent, shorter rides; older travelers are more inclined toward infrequent, longer rides. In another study, Alemi et al. (2018) model individuals' lifestyles using the California Millennials Dataset to identify the factors affecting the adoption of ride-hailing services. The results of this study indicate that highly-educated independent millennials who live in non-traditional households (living in core urban areas without owning personal vehicles) and without children have the highest adoption rate. In terms of personality traits, individuals with variety seeking and technology embracing attitudes are more likely to use ridehailing services.

Geographic context and built environment factors also play an important role in determining the usage frequency of ridehailing. According to Clewlow and Mishra (2017), the adoption of these services is comparatively higher in urban neighborhoods (29%) than those of suburban areas (15%). Alemi et al. (2018) also support a positive association between the demand for these services and the urbanization of the neighborhood. Along the same lines, the study by the Pew Research Center finds that a majority of users of these services are urban dwellers (21%), followed by suburban dwellers (15%) (Smith 2016). In a recent study, Yu and Peng (2019) investigate the relationship between different dimensions of the built environment and ridehailing demand using the 2016-2017 trip data from RideAustin, a local TNC company. The results support the findings of previous studies and also indicate that population density is a good predictor of ridehailing demand. Moreover, areas with higher roads and sidewalk densities are more likely to have a higher demand for TNC services.

TNC services are likely to reduce household vehicle ownership, but the extent of that reduction is not clear. The American Public Transportation Association (2016) reports that ridehailing users are more likely to own fewer cars. Similarly, Conway et al. (2018) use National Household Travel Survey (NHTS) data to examine the extent of the expansion of ridehailing services within the US and conclude that ridehailing users are more likely to be multimodal, owning fewer cars and using alternative modes of transportation. The results of a survey in Austin, Texas by Hampshire et al. (2017) support this relationship as these authors find that 9% of the ridehailing users purchased a vehicle after a suspension of these services, and 41% of users went back to driving.

The impact of TNC services on public transit is also unclear. Sadowsky and Nelson (2017) implement a regression discontinuity design to measure the effect of TNCs on public transit across 28 major urban US cities. The authors find that the introduction of Uber led to an increase in the use of public transportation, however, the introduction of Lyft after a few months had a negative impact on transit ridership. Sadowsky and Nelson hypothesize that the competition between these TNCs led to a decrease in cost and wait time, making these services more attractive than transit. Dias et al. (2018) analyze around one million trips by RideAustin and find that individuals living in neighborhoods with lesser access to transit have higher inclination to use ridehailing services, but there is a synergy between transit and ridehailing services in other areas. In another study, Barbar et al (2017) use a difference-in-difference design to quantify the impact of ridehailing services and observe a significant decrease in road-based public transit service, especially in areas with poor transit coverage, but an increase in the demand of subway and commuter rail. Hall et al. (2018) also adopt a difference-in-difference design and find that Uber is, on average, complementary to transit. However, the more detailed analysis shows a negative impact of Uber on the transit ridership in smaller cities, but a positive impact in larger cities.

Although no previous study has focused on eliciting preferences of TNC drivers, a few studies have investigated the demographics of such drivers, and the relationship between their desire to drive and personal attitudes. Hall and Kruger (2018) analyze data from two surveys in the US that were conducted in December of 2014 (N=601 drivers) and November of 2015 (N=632 drivers). Hall and Kruger find that 30% of the Uber drivers were between the ages of 30 and 39, 47.7% had college or advanced degrees, and only 14% were women. The authors also observe that more Uber drivers were single than married, and the married drivers had children at home. Berliner and Tal (2018) estimate the willingness of an individual to drive for TNCs using the stated preference data collected in Irvine, California in 2015. They find that the opportunity to make extra money followed by a fondness toward driving are key motivating factors behind an individual's willingness to drive for ridehailing services. Berliner and Tal also conclude that age, number of children, vehicle ownership, gender, and positive attitudes toward ridehailing have a significant role in estimating willingness to become a TNC driver.

This current research builds upon prior evidence on understanding the preferences of travelers and drivers for TNC services using revealed-preference, representative data of the TNC served areas in the United States. First, we expand the literature on understanding the socio-demographic characteristics of TNC users and drivers. Second, we help in identifying the demographic segments of TNC users who are interested in pooling rides. This question has received limited attention in the literature (Lavieri and Bhat, 2018), but it is crucial for policymakers to deploy pooling services faster and to make TNCs environmentally viable. Third, to further understand the impact of ridehailing services on greenhouse gas emissions, we recognize a group of TNC drivers who are willing to move toward fuel-efficient cars. Fourth, we also elicit preferences of TNC drivers to buy new vehicles with driving TNCs being a major purchase consideration, which have also not been explored in the literature.

## 3. Data and Summary Statistics

In this section, we provide details of the survey data and weight calculation, followed by a discussion on travel behavior and vehicle purchase decisions of TNC users, non-users, and drivers using data summary statistics.

### 3.1 Data Collection and Weight Computation
We use data from a survey of ride-hailing users, drivers, and non-users, which was conducted by Strategic Vision Inc. in 2017 among 11,902 U.S. consumers residing in TNC served areas. Figure 1 shows the geographical distribution of the number of respondents across the contiguous states of the USA. The procured survey data includes information on household characteristics and attitudes, as well as information on ride-hailing usage and preferences. Household characteristics include respondents' age, gender, marital status, education level, household income, ethnicity, residential location, mode of commuting, and number of members in the household.

The sample either under- or over-represents some demographic groups. For example, women above 49 living in metropolitan or urban areas with income less than $100,000 using public or non-motorized transport are under-represented in the sample and men above 54 living in small town areas with annual income greater than $100,000 are over-represented in the sample. To address this concern, we estimate person-level weights using the iterative proportional fitting (IPF) technique (Bergmann, 2011). IPF matches the joint probability distribution of various demographic characteristics in the collected sample and the population-level datasets on urban and rural housing units (2016 American Community Survey data and 2010 U.S. Census Bureau data). In other words, we compute weights by scaling the survey sample proportions, in four demographic classes or 32 categories (four gender- and age-based, two income-based, two travel mode-based, and two residence location-based groups), relative to the corresponding class-specific proportions in the population-level data. We implement this IPF method using the ipfweight package in Stata. The estimated weights vary between 0.15 - 4.65. All the results presented in this paper are based on the weighted sample.

**[Insert Figure 1 here]**

### 3.2 Explanatory and Response Variables

Table 1 summarizes statistics of the population-weighted explanatory variables used in all logistic regression models of this study. The sample statistics are consistent with those of the population. For example, in the weighted sample, the proportion of males, average age, and average annual income are 48%, 46.77 years and $94,909, respectively. These numbers for the population are 49%, 45.75 years and $81,346, respectively. Around 19% of respondents are early adopters of these ridehailing services. Among 1541 TNC drivers in the sample, around 25% drive daily.

Key summary statistics of the response variables are shown in Table 2. The results indicate that the sample proportions of TNC users, drivers, and non-users are 29%, 29%, and 42%, respectively. Among the ridehailing users, 13% of them had used ridepooling services in the past. Among drivers, 53% of them indicated that they would be inclined to shift towards more fuel-efficient vehicles. When asked about the change in vehicle ownership preferences of drivers, 47% of them had a high propensity towards buying or leasing a new vehicle as a result of driving for TNCs.

**[Insert Tables 1 and 2 here]**

### 3.3 Mobility Patterns of Ridehailing Users

We created a two-way table to understand the association between the most used mobility option by respondents and their TNC usage frequency (see Table 3). We label "frequent TNC users" to those who use TNCs once or more times per week, and the remaining respondents fall in the category of "infrequent TNC users". The results suggest that personal vehicles and ridehailing are the most used travel modes by around 53-61% and 22-32% frequent TNC users, respectively. However, these proportions are 79-87% and 3-5% for the infrequent TNC users, with a marginal decline in the share of transit among the most used travel modes (4-10% for frequent vs. 3-6% for infrequent users). This pattern indicates that ridehailing does not impact transit demand significantly, but rather personal vehicle users are mainly shifting to TNCs.

In Table 4, we analyze how the unavailability of the most used travel mode affect mobility patterns of travelers. Around 66% and 14% of those who mostly use ridehailing are likely to switch to driving personal vehicles and transit, respectively, in the absence of these services. Hampshire et al. (2017) also observed a similar trend where 41% of the riders went back to driving after the suspension of ridehailing services in Austin. Among those who indicated their most used mode to be driving personal vehicles, 31% and 46% would shift to ridehailing and carpooling/carsharing, respectively, in the absence of their preferred mode. These proportions are 29% and 4% for the frequent transit users.[2] All these findings further strengthen our earlier result that ridesharing and carsharing services are mainly capturing personal-driving demand, without substantially affecting demand for transit. In contrast to Clewlow and Mishra (2017), we find evidence against the impact of ridehailing service on induced travel demand – only 0.45% of frequent ridehailing users would not have made those trips if these services were not available. These discrepancies can be attributed to different target samples of both studies. The sample used by Clewlow and Mishra (2017) had an oversampling of respondents in San Francisco and Los Angeles.

We present other insightful statistics about preferences of ridehailing users. Around 10% of ridehailing users reported postponing the purchase of a new car. This result is similar to that of Hampshire et al. (2017), who found that 9% of the ridehailing users purchased a vehicle after the suspension of these services in Austin. In terms of trip purpose, a high proportion (around 46%) of respondents selected ridehailing, carsharing or carpooling as their mobility options on trips to/from social events. This finding is consistent with the previous studies, which found the most common usage of ridehailing on trips for recreational activities (Alemi et al., 2018; Lavieri and Bhat, 2018; Young and Farber 2019). The above finding is further supported by a sample statistic that around 21% of ridehailing users reported "didn't want to drive after drinking" as the most important reason for using ridehailing services. "Convenience" was reported as the other most important reason behind using ridehailing service by around 24% of respondents. These results are aligned with those of previous studies, which also found these two factors to be the main reasons for preferring ridehailing (Rayle et. al. 2016, Alemi et. al. 2018, Conway et. al. 2018, Young and Farber 2019).

We further analyze relatively small proportion, 13%, of ridehailing users who have used ridepooling services. Around 34% of ridehailing trips were pooled by these ridepooling users in the past. When asked about the reasons for not using ridepooling services, a large percentage of the ridehailing users, around 50%, mentioned that they had not heard of these services. Another dominant reason, which was indicated by around 22% of the ridehailing users, is a preference for "private rides".

**[Insert Tables 3 and 4 here]**

### 3.4 Preferences of non-users
We now discuss reasons of not using ridehailing by non-users. The highest proportion (around 36%) of the respondents mentioned that they would just prefer driving by themselves. Among the other reasons, around 21% of the nonusers did not need taxi or ridehailing services in the past. Non-users of TNCs, who use personal vehicles for first/last-mile transit connectivity, were asked about their preferences to switch to TNC services for the last-mile. Approximately 41.3% of these non-users reported their willingness to make that switch.

---

[2] However, proportion of shift to ridehailing is similar for both personal vehicle and transit users, in absence of both services, personal vehicle users would mainly contribute to ridehailing market share due to their much higher current share (86% in the sample).

### 3.5 Preference of TNC Drivers

Drivers were asked about their emotional response to driving for TNCs. Whereas around 54% of the drivers reported their experience to be excellent, 28% and 18% felt neutral and unsatisfactory, respectively. As expected, TNC drivers who work more frequently end up driving more miles per week. Specifically, the drivers working daily, and every other day drive 42 miles and 31 miles per week, on average, respectively, but those who work less than once per month, only drive an average of 16 miles per week.

We created a two-way table to understand the relationship between the decision to rent/lease/purchase a vehicle and their driving frequency (see Table 5). Around 65% of drivers who work daily indicated that driving for TNCs was a consideration in terms of new vehicle ownership. As expected, this proportion declines to 51% and 40-46% for those who drive every other day and once a week or less, respectively. Moreover, around 93% of drivers use their primary vehicle to drive for ridehailing services. This proportion remains intact across different driving frequencies of drivers.

We then asked about downtime activities of drivers. Around 29% of drivers reported that they drive to the busy parts of the city to get more rides, inducing extra vehicle miles traveled (VMT). Average pick-up time during peak and off-peak hours are 9 and 10 minutes, respectively, which further adds around 2-3 miles per trip to the induced VMT. In terms of inclination toward future fuel type, 26% of TNC drivers who work more than 20 hours per week would prefer Diesel, but this proportion is 8-11% for those who drive less frequently. Around 25% of drivers would like to drive a hybrid electric vehicle in the future.

**[Insert Table 5 here]**

## 4. Results and Discussion

Apart from parameter estimates of the link function of the logit models, we also report relative risk or odds-ratio estimates with 95% confidence intervals. We explore non-linear effects of continuous variables by plotting the predicted choice probabilities and marginal effect estimates over the support of the covariate. To compute these quantities at a given level of the covariate of interest, we fix all other covariates at their sample mean and make a unit change in the specific covariate of interest.

### 4.1 Model 1 – Preference of being a TNC Rider or a Driver

Table 6 shows the parameter estimates and the relative risk ratios of the multinomial logistic model, which explains the association between individuals' sociodemographic characteristics and their preference for being a TNC rider, driver, or non-user.

The predicted choice probabilities and the marginal effects plots of age (see Figure 2) indicate that the preference for being a TNC driver and non-user, respectively decreases and increases with an increase in age, keeping all other characteristics the same. This result is consistent with the findings of Hall and Krueger (2018) who also had a higher proportion of younger people among Uber drivers. Perhaps, platforms such as Uber provide new opportunities to the younger population who are more open to a flexible work schedule and are willing to take up multiple jobs. However, we observe a non-linear effect of age on the propensity of being a TNC user. The likelihood of being a user increases with the increase in age for someone younger than 44 years, but the pattern is reversed post 44 years. This change in direction of the marginal effect of age at 44 years is insightful as the previous studies have reported a linear and negative association between age and propensity of being a TNC user (Kooti et al., 2017; Alemi et al 2018). Finally, the higher inclination of younger people toward being a TNC user or driver is also aligned with the education and psychology literature where previous studies have established that younger people are more likely to adopt information and communication technologies (Helsper and Eyon, 2010; Milojev and Sibley, 2017).

Higher income individuals have a higher inclination to ride TNC services, but a lower propensity to become a TNC driver. These results are consistent across all income values (see Figure 3) and are also aligned with previous studies (Rayes et al 2014; Clewlow and Mishra 2017). A probable explanation of this relationship is higher affordability of wealthier individuals to ride TNC services and at the same time, lesser likelihood to become a TNC driver for additional income. After controlling for key covariates such as education, age, marital status, and gender, individuals with a higher household vehicle ownership are less likely to use TNCs in any form (driver or rider). However, variation in the predicted choice probabilities due to a change in vehicle ownership is very small (see relatively flat predicted choice probability plots in Figure 4 indicate). This relationship is intuitive because TNC services in households with high vehicle ownership might serve as a convenient travel option rather than a frequent mode of transport. On the same line, Bhat and Lavieri (2018) also report a decrease in ride-hailing frequency with the increase in vehicle availability.

Moreover, early adopters of technologies and residents of metropolitan areas are more inclined toward riding and driving ridehailing services, ceteris paribus. In fact, magnitudes of relative risk ratios indicate that these two covariates have the most practically significant relationship with the individual's preference to ride or drive TNC services. Being an early adopter and being a metropolitan resident increases the odds of being a driver (cf. being a nonuser) by factors of 4.81 and 1.94, respectively, and increases the odds of being a rider by factors of 1.36 and 1.53, respectively. These findings are consistent with the literature. For instance, Alemi et al. (2018) find that early adopters of TNC services are likely to be "technology-oriented" and thus they tend to adopt such services in bundle as a part of their modern lifestyle. Similarly, Hall and Krueger (2018) and Alemi et al. (2018) also observe higher inclination of metropolitan residents toward using these services. Lavieri and Bhat (2018) speculate three possible reasons for this tendency, namely parking restrictions in urban areas, lower trip costs due to shorter trip lengths, and higher reliability of TNC services. In fact, higher travel demand in metropolitan areas also explains residents' higher propensity of being a driver.

Finally, postgraduate degree holders and single individuals, everything else constant, are more likely to ride TNC services but are less likely to be a driver. These results are aligned with the findings of Lavieri and Bhat (2018).

**[Insert Table 6 here]**

**[Insert Figures 2 to 4 here]**

## 4.2 Model 2 – Preference to Ridepool

Table 7 shows the parameter estimates and the odds ratios of a binary logistic model which explain the preferences of TNC riders for ridepooling (base category: non-users of ridepooling). As the literature on ridepooling is still at a nascent stage, we have highlighted some of the similarities of our results with the results of carpooling/carsharing studies.

The predicted probability and marginal effect plots of age (see Figure 5) indicate that the tendency of pooling consistently decreases with the increase in the age of TNC users. Lavieri and Bhat (2018) observe a similar trend, which they attribute to tech-savviness and variety-seeking behavior of the younger population. Carpooling is also more common among individuals between the ages of 25 and 55 (Shaheen et al., 2016). However, the relationship between age and propensity to ridepool varies with sociodemographic characteristics such as education level (see Figure 6) and gender (see Figure 7). Clearly, the inclination of a male (and postgraduate) TNC user to pool is less severely affected by age as compared to that of females (and below postgrad), keeping all other characteristics the same. A young female TNC user is much more likely to pool than that of a young male. More specifically, a female TNC user who is younger than 54 years has a higher probability of pooling than a male user of the same age, but the trend reverses for users older than 54 years. The transition age for education effect is around 34 years. In other words, TNC user with education below postgraduation, who is younger than 34 years, is likely to have higher tendency to use pooling than a postgraduate TNC user of the same age, but the pattern is reversed for a TNC user older than 34 years. We thus provide new insights on the non-linear relationship between TNC riders' preference for ridepooling and their gender and education levels.

Household vehicle ownership is negatively associated with the preference of a TNC user to ridepool: odds of ridepooling by a TNC user due to the addition of a household vehicle decreases by a factor of 0.87. This relationship is consistent with the findings of Lee et al. (2018) in the context of carpooling. The predicted probability plot in Figure 8 shows the linear nature of this relationship over the support of vehicle ownership.

Similar to the results of Model 1 in section 4.1, whereas metropolitan resident and early adopters are more inclined, males are less inclined to ridepool as compared to their counterparts, ceteris paribus. A few studies in the environmental psychology literature (e.g., Glover et al. 1997) argue that females take a stronger standpoint on ethical, environmental and pro-social behavior as compared to males, which can be a plausible reason behind a higher inclination of females toward ridepooling. Residential location is the most practically significant predictor as reflected in odds ratios – living in metropolitan areas increase the odds of a TNC user to ridepool by a factor of 1.73 as compared to those living in suburban areas. This result is consistent with the findings of Almei et al (2018) and Lavieri and Bhat (2018).

**[Insert Table 7 here]**

**[Insert Figures 5 to 8 here]**

## 4.3 Model 3 – Preferences of TNC Drivers for Fuel-efficient Vehicles

Table 8 summarizes results of the binary logistic model that identifies the characteristics of TNC drivers who would prefer to switch to fuel-efficient vehicles (base category: prefer not to switch to fuel-efficient vehicles). Since there is no study on such preferences of TNC drivers, we highlight some similarities and differences between our results and the results of studies concerned with characteristics of electric vehicle (EV) buyers.

As expected, the predicted probability plot shows a negative correlation between the age of TNC drivers and their propensity to switch to fuel-efficient vehicles (see Figure 9). Hidrue et al. (2011) observe a similar trend among EV buyers. The relationships between the inclination of a driver to switch to fuel-efficient vehicles and the driver's residential location (see Figure 10), education attainment (see Figure 11), and marital status (see Figure 12) vary across age of the driver.

More specifically, postgraduate drivers who live in metropolitan areas are more prone to fuel-efficiency than their counterparts if their age is below 48 years, ceteris paribus. This pattern reverses for drivers older than 48 years because the inclination of these educated metropolitan residents toward fuel-efficient vehicles decreases more rapidly than their counterparts with the increase in their age. These results are consistent with the previous studies, which have found that younger and highly educated individuals are more prone to buy alternative fuel vehicles (Dütschke et al. 2013, Hackbarth and Madlener, 2013). As a matter of fact, being a postgrad and metropolitan resident are the two most significant predictors of propensity to switch to a more fuel-efficient vehicle. The individuals with higher education level have more awareness of the environmental impacts of automobiles; and being a metropolitan resident is associated with having a variety-seeking lifestyle (Franzen and Volg, 2013; Laiveri and Bhat, 2018).

In addition, married drivers are more inclined to switch to fuel-efficient vehicles than their single counterparts for the age below 60 years. These results are consistent with those of Peter et al. (2011), who find that households with children are more likely to be EV buyers. In the context of this study, married drivers are probably even more conscious about fuel-efficiency of the vehicle because they might end up driving more miles than their single counterparts due to the use of the same vehicle for other household activities. A recent report by Ipsos (2017) also ascertains that married households are more inclined toward buying electric vehicles.

Drivers who are early adopters of ridehailing services are found to be more inclined to switch to fuel-efficient vehicles. In fact, being an early adopter increases the odds of a driver to switch to a fuel-efficient vehicle by a factor of 1.47. This observation makes intuitive sense because early adopters are likely to have a higher orientation toward technology and thus might be more interested in driving technologically advanced vehicles. The results of the survey conducted by CleanTechnica are also consistent with our findings where 38% of the respondents selected "love for new technology" as the reason for switching to the fuel-efficient vehicles.

TNC drivers who drive daily are more inclined to use fuel-efficient vehicles. Perhaps, these regular drives are more sensitive to fuel price and can foresee the benefits of investing in fuel-efficient vehicles due to their higher vehicle-miles-traveled. It is well established in the EV literature that early adopters of EVs drive a higher number of kilometers (Plotz et al. 2014). In fact, an individual's willingness to pay for EVs is also primarily driven by savings in fuel costs, which is further associated with vehicle-miles-traveled (Hidrue et al., 2011).

Drivers with higher vehicle ownership also have a higher tendency to switch to fuel-efficient vehicles. The probability of switching linearly increases with the increase in the number of vehicles (see Figure 13). Since the income of drivers is not controlled in the specification (because it was not statistically significant), perhaps the missing income effect is also reinforcing pro-fuel-efficiency behavior of drivers with high vehicle ownership. We observe similar relationships in the EV literature. Previous studies have shown that higher-income consumers are more likely to buy EVs (Erdem et al., 2010; Saarenpaa et al., 2013). Moreover, EVs are typically owned by high-income households with more than one car (Hjorthol, 2013).

**[Insert Table 8 here]**

**[Insert Figures 9 to 13 here]**

### 4.4. Model 4 – Preference of TNC Drivers to Buy a New Vehicle

We use a binary logistic regression to estimate whether driving for a ridehailing service was a consideration in TNC drivers' decision to buy, rent or lease a new vehicle. Results are shown in Table 9 (base category: no change in the preference to buy a new vehicle as a result of driving for TNCs).

The predicted probability and marginal effect plots (in Figure 14), respectively, indicate that the tendency to buy a new vehicle with driving for TNCs being a major consideration decreases with the increase in age of the TNC driver, keeping everything else constant. For a large support of the driver's age (below 55 years), postgraduate and married drivers have higher propensity than their counterparts, but the pattern gets reversed for the older drivers (see Figures 15 and 16) because the negative effect of the increase in age is much higher for these drivers than their counterparts. Married drivers have higher incentives to buy an additional vehicle, perhaps because the additional vehicle can serve as the second vehicle for other household activities.

Higher income drivers have a lower inclination to buy a vehicle with driving for TNCs as a consideration (see predicated probability and marginal effect in Figure 17). Further investigation of this income effect within single and married drivers indicates that the increase in income of married drivers increases their probability, but the reverse effect is seen for single drivers (see Figure 18). Overall, the negative income effect is a manifestation of the steeper rate of decline in the probability for single drivers as compared to the rate of increase in the probability for married drivers. For example, the predicted probability is the same (around 0.48) for a single and a married driver when the annual income is below $10,000, but this probability increases to 0.51 for a married driver and decreases to 0.38 for a single drive if the driver's income increases to $100,000, keeping all other characteristics the same. The high propensity of low-income single drivers to buy, lease, or rent a vehicle with driving for TNCs being a major consideration in part explains why some of TNCs offer such renting and leasing services to attract drivers from this demographic group.

A male driver residing in metropolitan areas, who is an early adopter, has higher vehicle ownership, and drives daily, is likely to have a higher inclination to buy a vehicle with driving for TNCs being a major purchase consideration than the counterpart, ceteris paribus. In terms of the strength of these relationships, being an early adopter of technologies increases the odds of changing the preference of a driver by a factor of 2.59, driving daily by a factor of 2.09, and being a metropolitan resident increases the odds by a factor of 1.90.

**[Insert Table 9 here]**

**[Insert Figures 14 to 18 here]**

## 5. Conclusions and Future Research

This study adds new insights to the existing literature on understanding preferences to use ridehailing services by identifying relationships of individuals' socio-demographic characteristics with their preferences to use ridehailing services (as a driver or a rider), and the willingness to ridepool by users. We have accomplished these findings by calibrating multinomial and binary logistic models using a unique dataset involving survey of respondents (N=11,902) in TNC served US cities in 2017. The uniqueness of this study stems from estimating preferences of TNC drivers to buy new vehicles with driving for TNCs being an influencing purchase consideration and to shift to fuel-efficient vehicles, which have not been explored in the literature. We also observe non-linear relationships by accounting for interaction effects of continuous covariates (e.g., age and income) with binary covariates (e.g., gender) while estimating preferences.

Results indicate that younger individuals who have achieved higher education levels, live in metropolitan areas, and belong to more affluent families are more likely to use ridehailing services. However, the relationship between the individuals' probability to use TNC services and their age is downward parabolic: it increases until the age of 48 years and then decreases. Households with higher vehicle ownership are less likely to associate themselves with TNCs in any form (as a driver or a rider). Further exploring the inclination of the ridehailing users to ridepool, we find that older travelers with higher household vehicle ownership who are living in suburban areas are less likely to pool rides. In terms of interaction effects, females with an education level below postgraduation are more likely to ridepool than their counterparts if they are younger than 34 years, but the pattern gets reversed among travelers older than 54 years.

We now discuss key insights about the preferences of TNC drivers. Younger and married drivers who drive daily and own a higher number of vehicles are more likely to switch to fuel-efficient vehicles, ceteris paribus. These results are consistent with previous studies eliciting preferences for electric vehicles. Further, interaction effect estimates reveal that postgraduate drivers who live in metropolitan areas are more pro-fuel-efficiency than their counterparts if their age is below 48 years. Finally, the tendency of married TNC drivers to buy a new vehicle with driving for TNCs being a major consideration increases with income.

These findings can inform different stakeholders (such transportation planners, government agencies, automakers, and TNCs) in developing policies to encourage ridepooling and deployment of high fuel economy vehicles, which can further help in realizing system-level environmental benefits of these services. For instance, automotive manufacturers and auto-leasing companies could partner with TNCs to offer attractive leasing plans for the identified pro-fuel-efficiency drivers to encourage adoption of high fuel economy vehicles including electric vehicles. Such a partnership could be mutually beneficial for all parties. On one hand, it could help automakers meet their respective zero emission vehicle (ZEV) mandate targets in the ZEV states as well as their federal fleet fuel economy targets. On the other hand, it would also benefit the high mileage TNC drivers by helping them reduce their operating fuel costs. Finally, deployment of a higher number of fuel-efficient vehicles would allow TNCs to balance supply and demand for new environment-friendly initiatives (e.g., "green mode" initiative by Lyft (Price, 2019)), which allow riders to specifically call green vehicles. The attractive lease options offered by GM on Chevrolet Bolt for Uber and Lyft drivers is a case-in-point (Kurczewski, 2017).

A targeted campaign can be organized to spread awareness about the environmental benefits of ridepooling to encourage ridehailing users to pool rides. This is especially true as our research showed that younger female passengers, who are known to be more environmentally conscious, prefer ridepooling. To ensure passenger safety while pooling, automakers could be encouraged to provide tailor-made vehicles for pooling that include partitions.

Readers should be careful in interpreting results as this study does not identify causal relationships between individuals' socio-demographic characteristics and their preferences, instead we only explore correlation patterns. Collaborations with TNCs to conduct randomized experiments in order to disentangle causal effects can be a potential avenue for future research.

# Figures

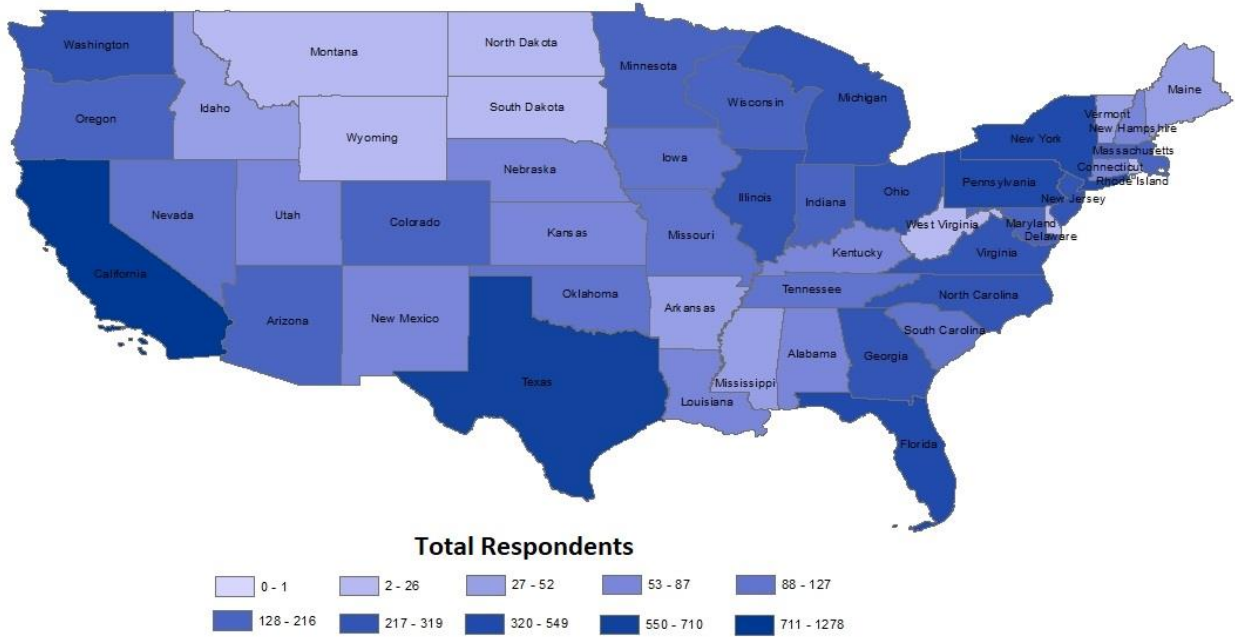

**Figure 1:** Geographical distribution of "the number of respondents" at state-level (N=11,902)

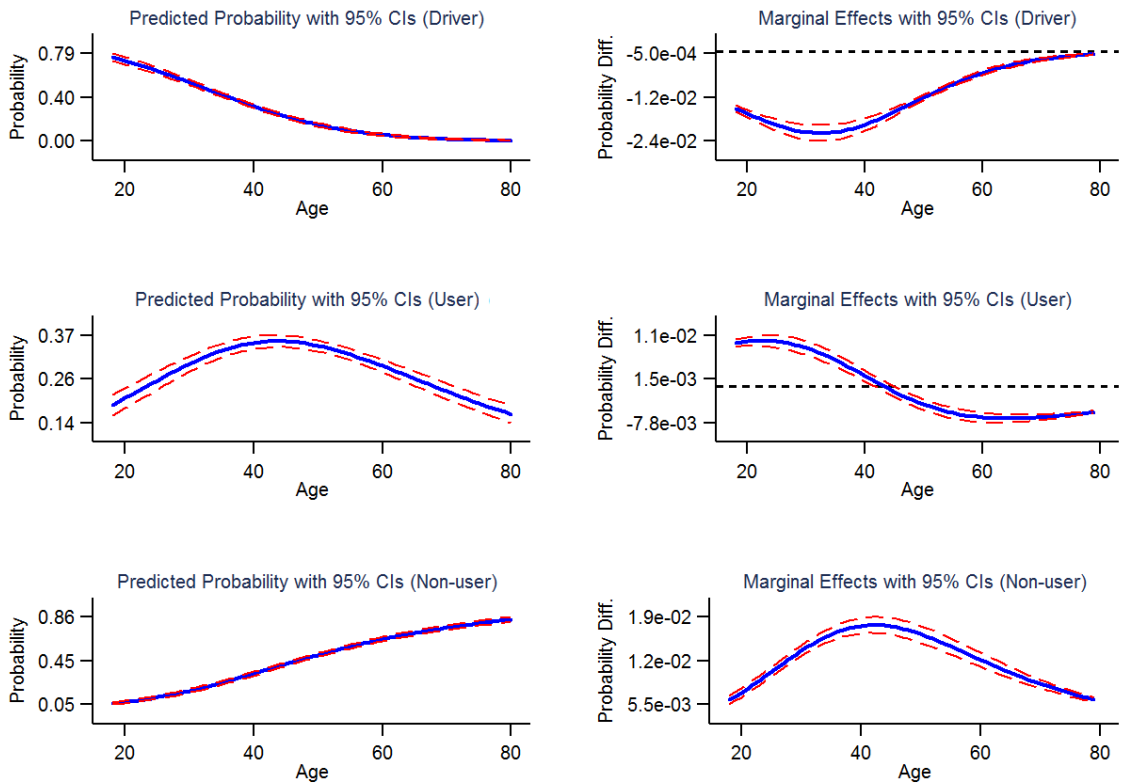

**Figure 2:** Predicted Probability and Marginal Effect of "Age" (Model 1).

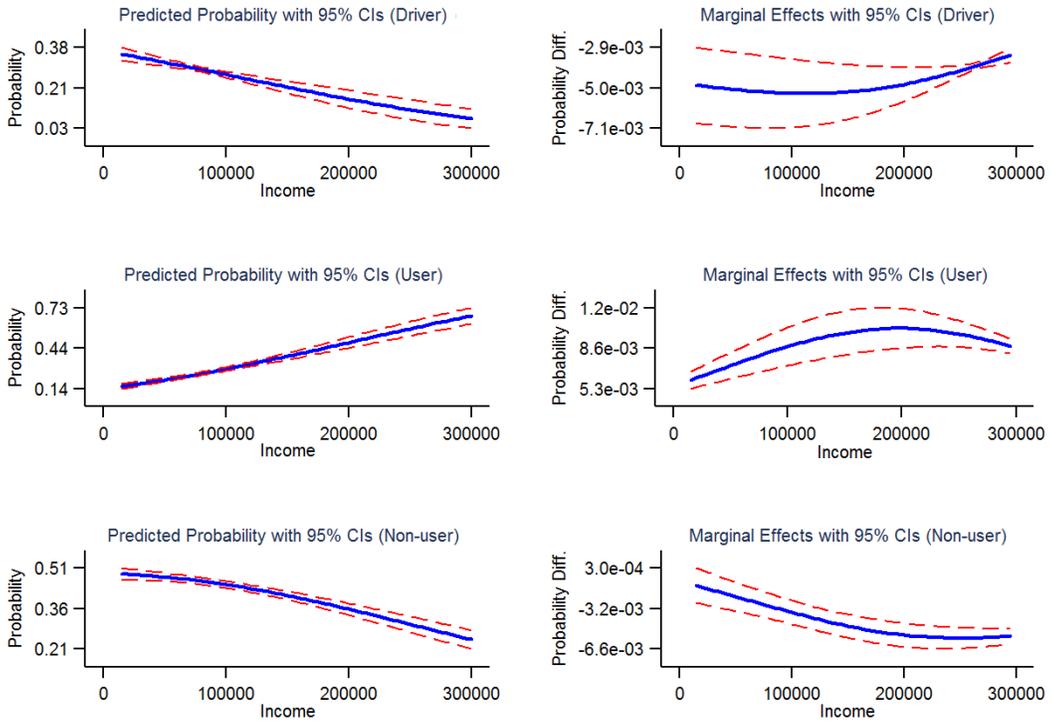

**Figure 3:** Predicted Probability and Marginal Effect of "Annual Income" (Model 1).

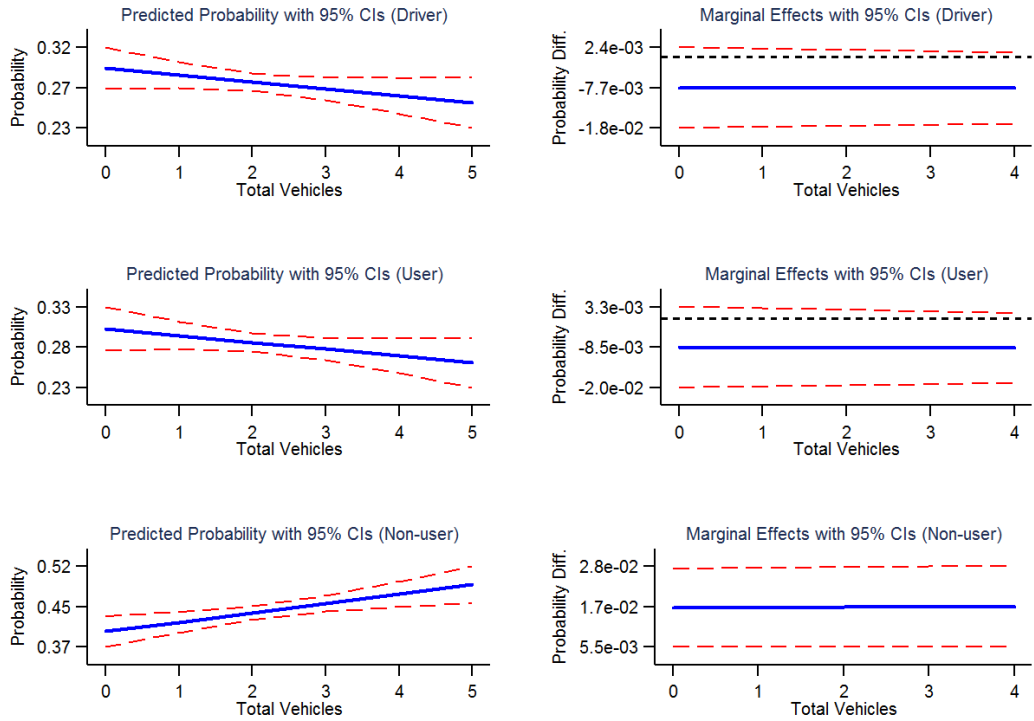

**Figure 4:** Predicted Probability and Marginal Effect of "Total Vehicle Ownership" (Model 1).

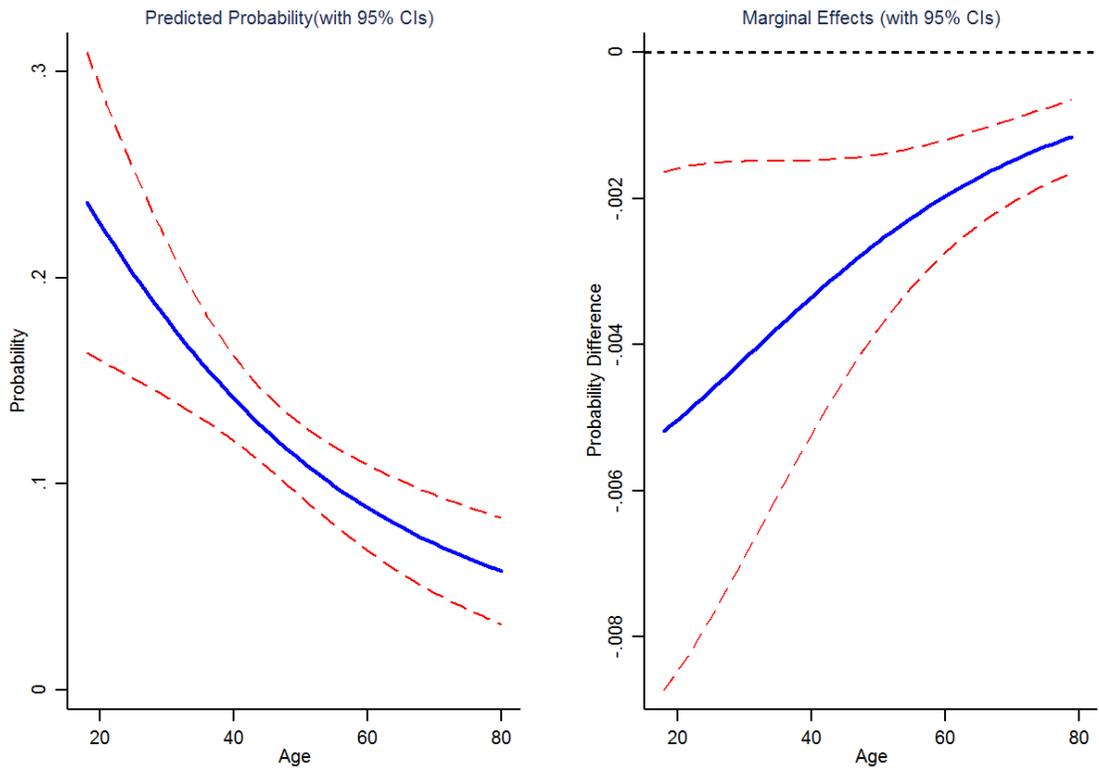

**Figure 5:** Predicted Probability and Marginal Effect of "Age" (Model 2).

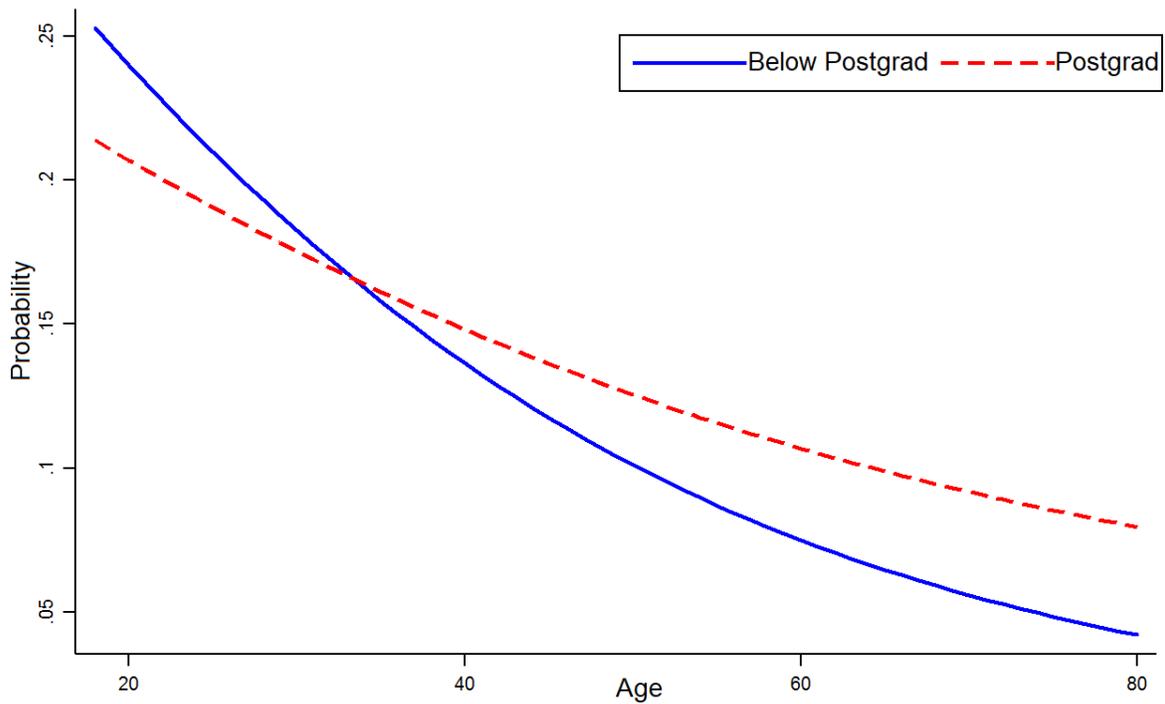

**Figure 6:** Interaction Effect of "Age" and "Postgraduation indicator" (Model 2).

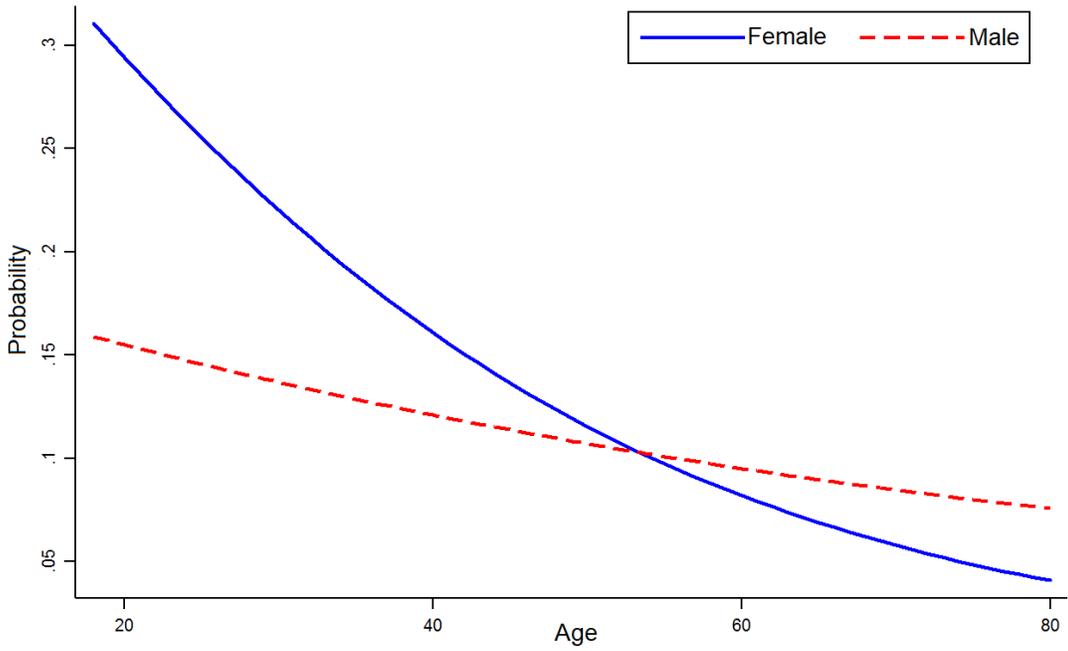

**Figure 7:** Interaction Effect of "Age" and "Male indicator" (Model 2).

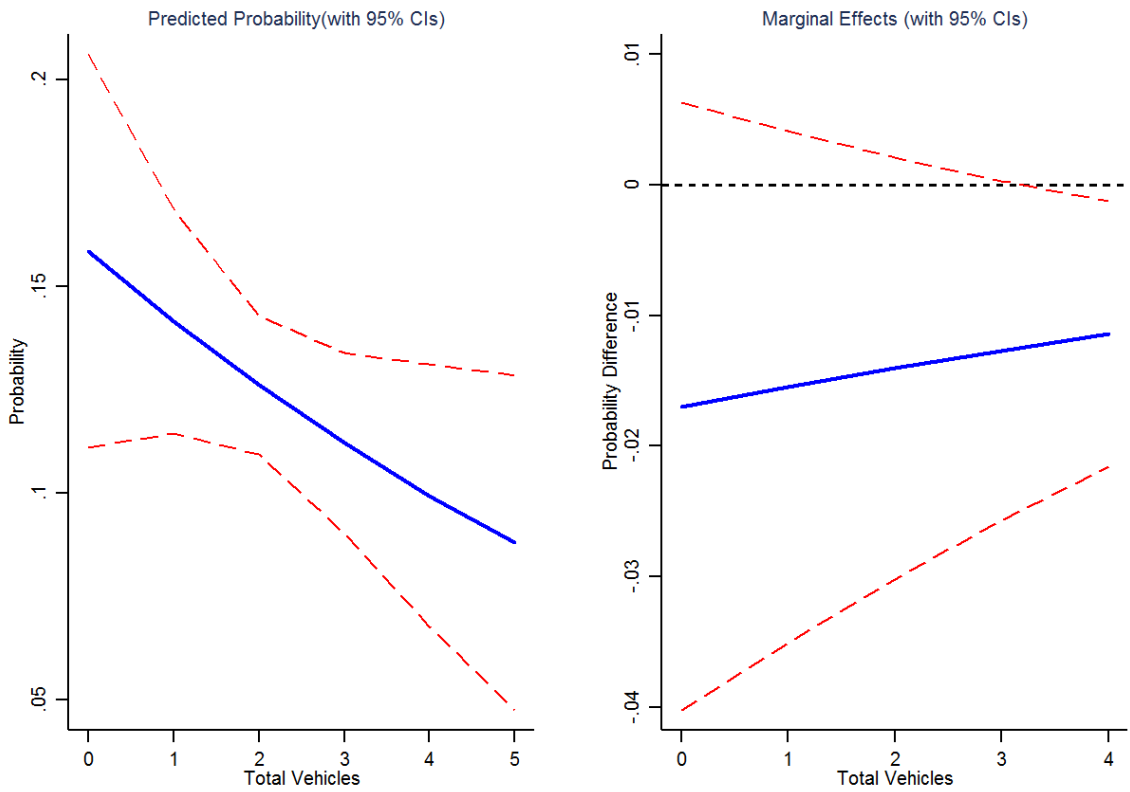

**Figure 8:** Predicted Probability and Marginal Effect of "Total Vehicle Ownership" (Model 2).

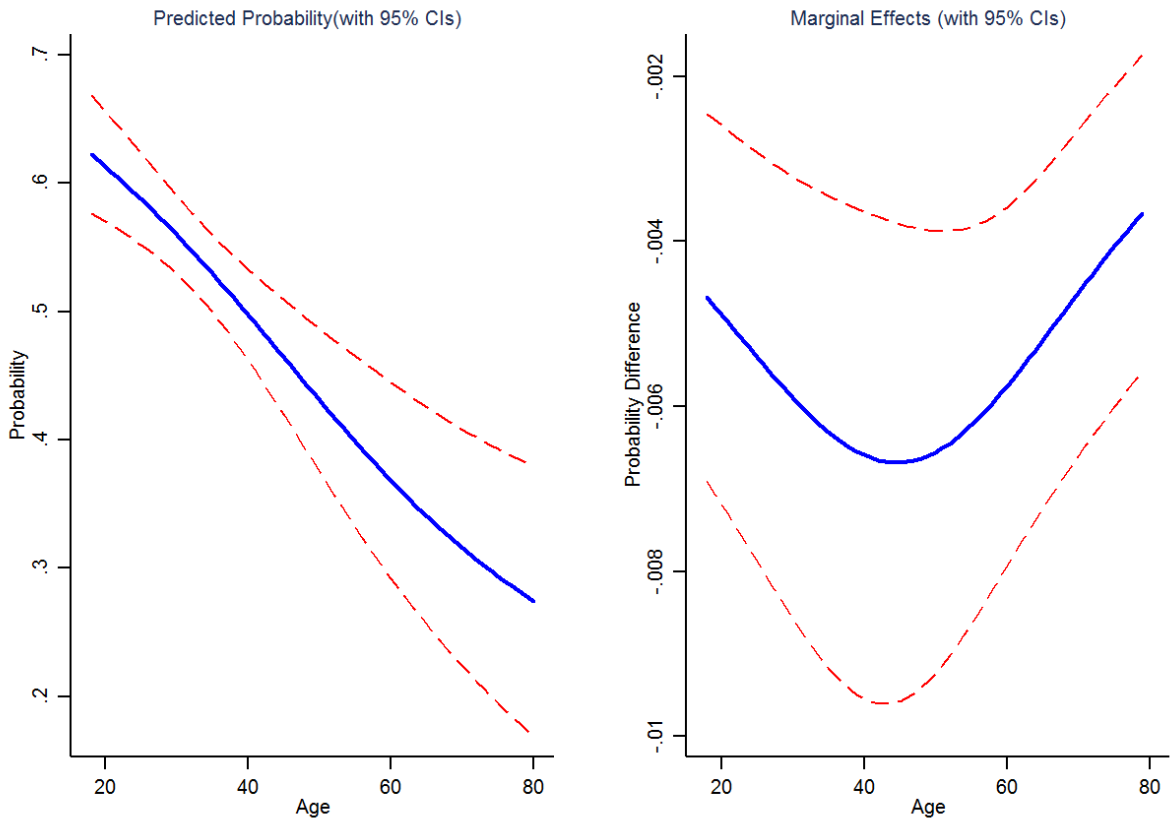

**Figure 9:** Predicted Probability and Marginal Effect of "Age" (Model 3).

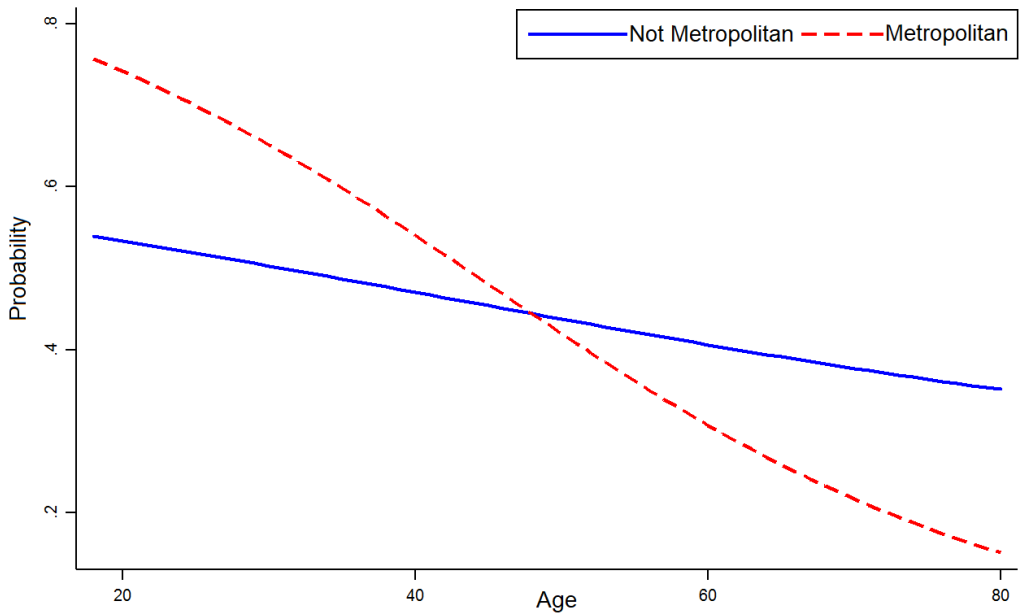

**Figure 10:** Interaction Effect of "Age" and "Metropolitan Resident indicator" (Model 3).

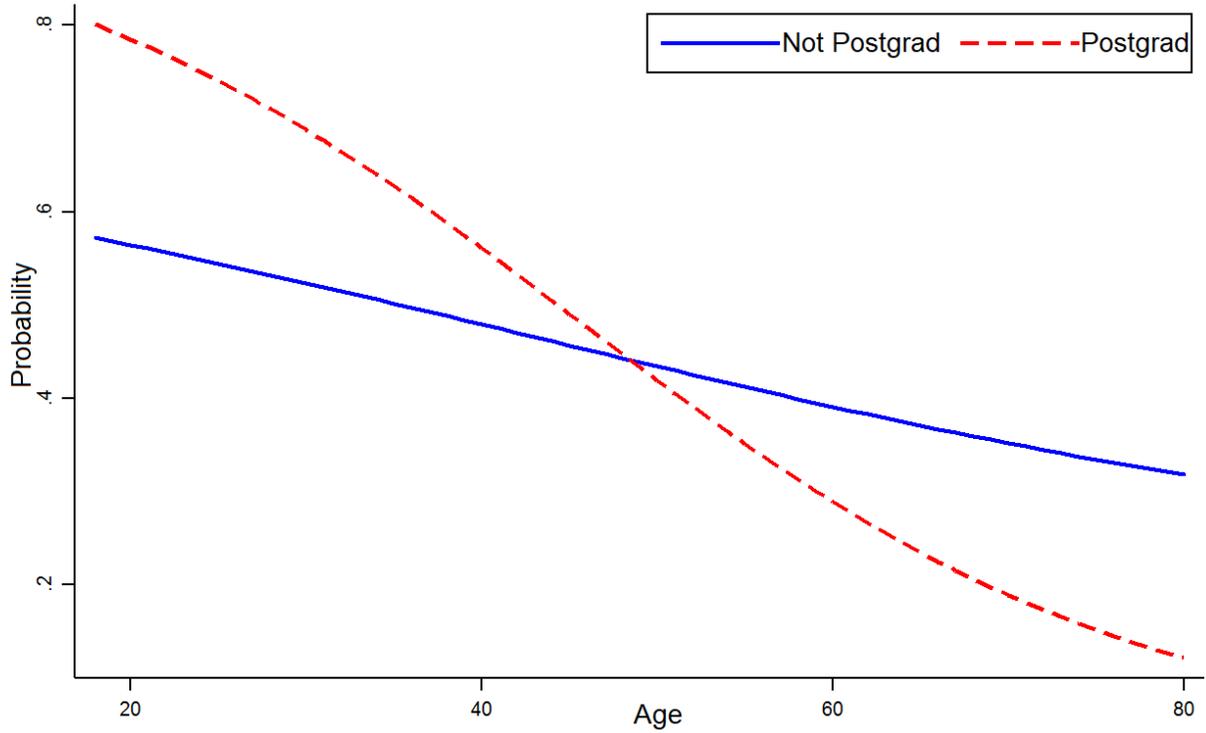

**Figure 11:** Interaction Effect of "Age" and "Postgraduation indicator" (Model 3).

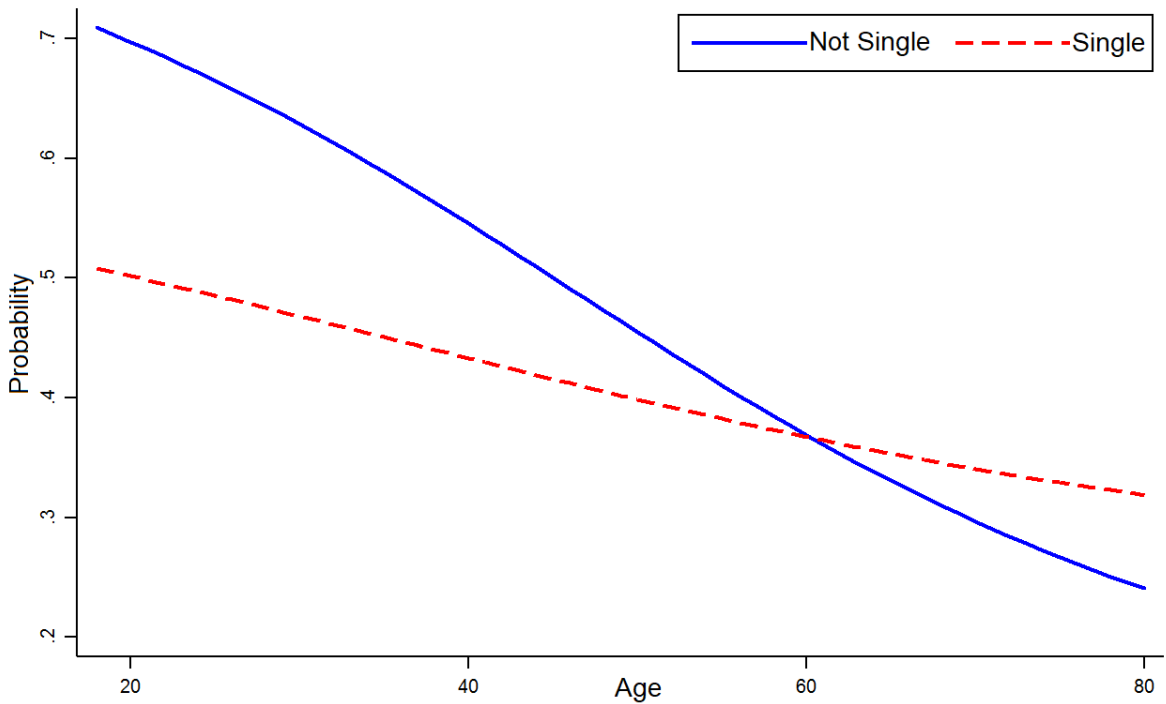

**Figure 12:** Interaction Effect of "Age" and "Single indicator" (Model 3).

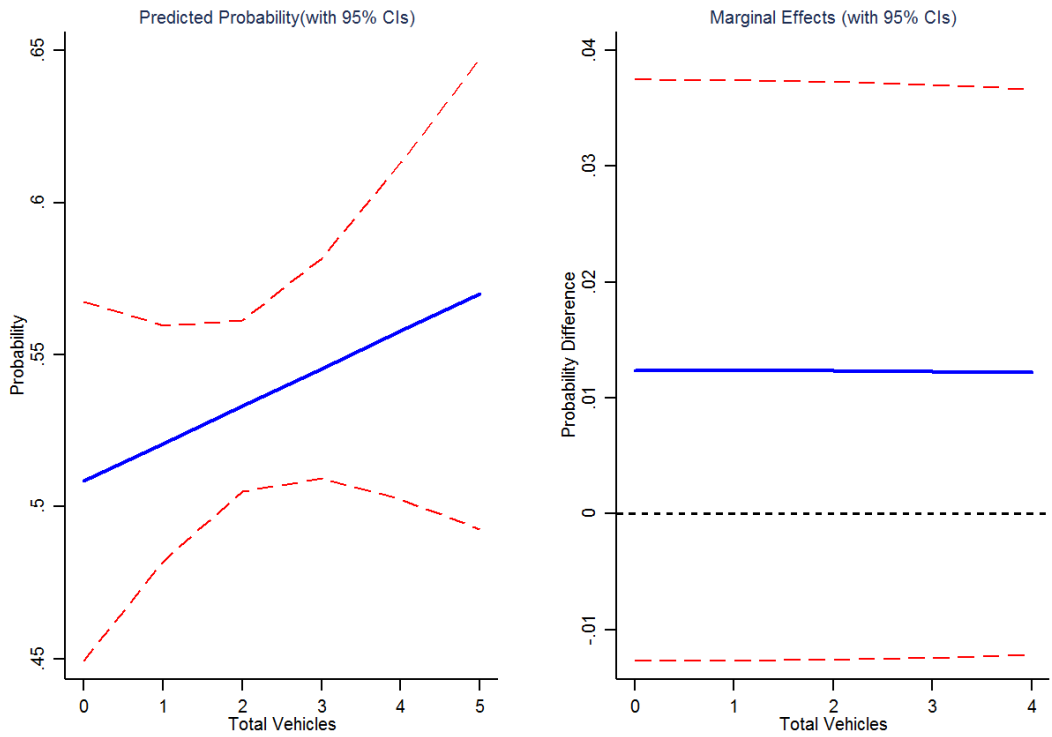

**Figure 13:** Predicted Probability and Marginal Effect of "Total Vehicle Ownership" (Model 3).

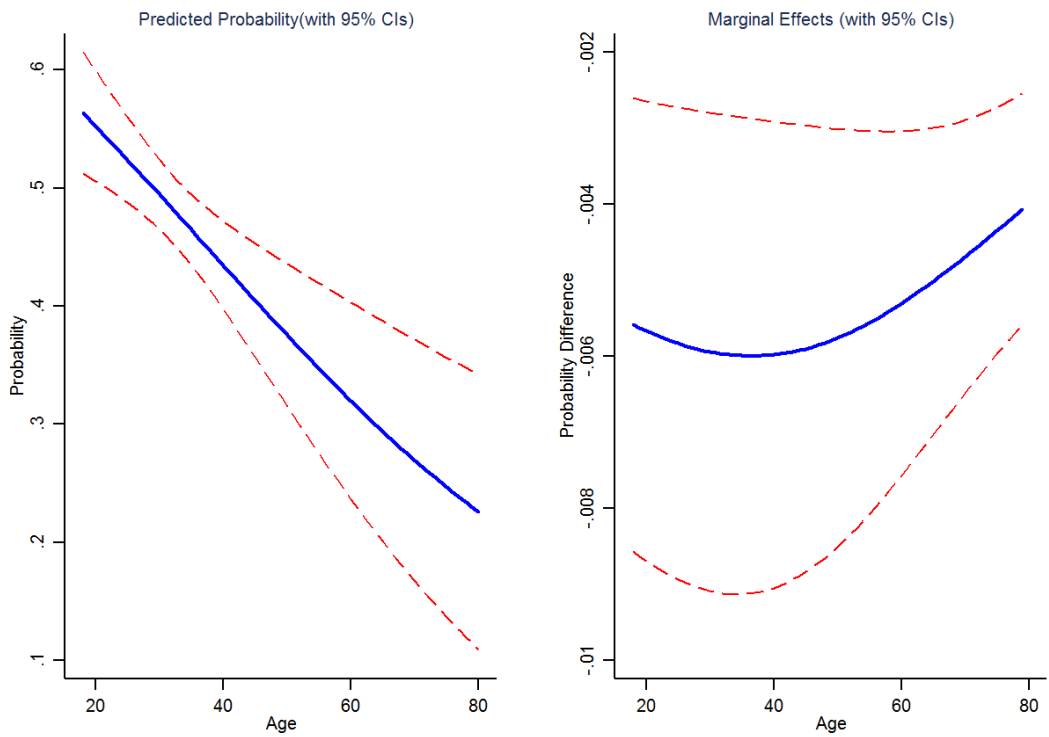

**Figure 14:** Predicted Probability and Marginal Effect of "Age" (Model 4).

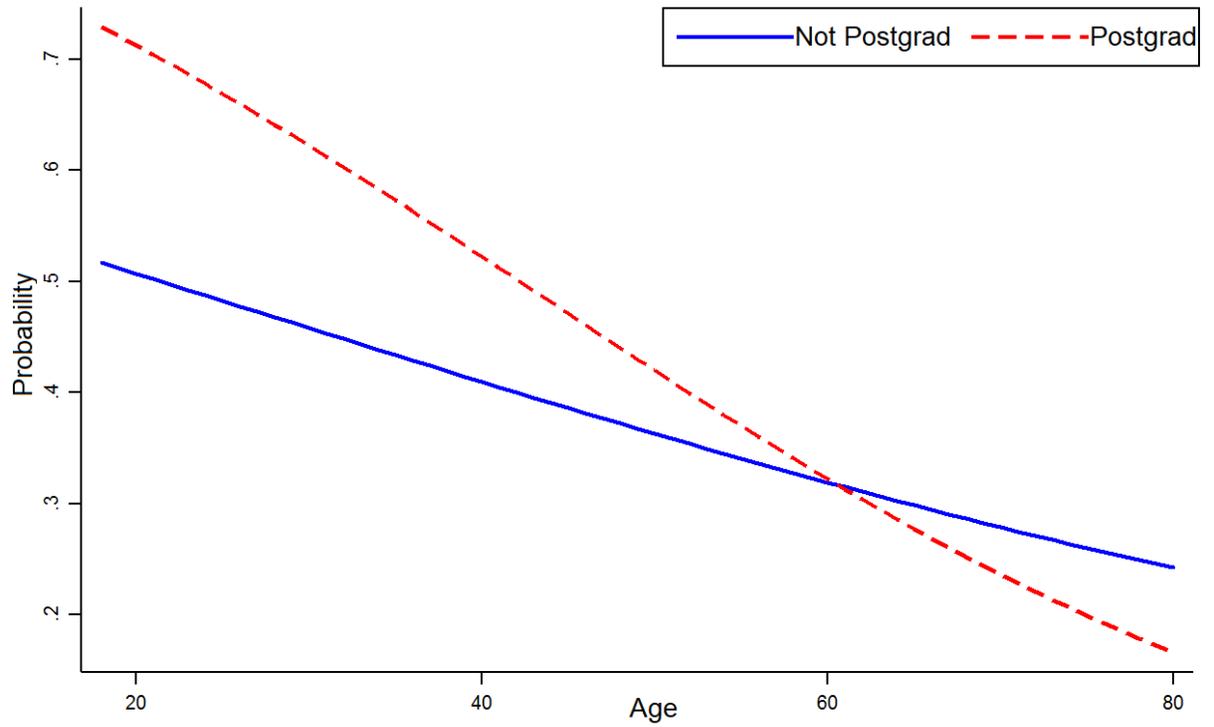

**Figure 15:** Interaction Effect of "Age" and "Postgraduation indicator" (Model 4).

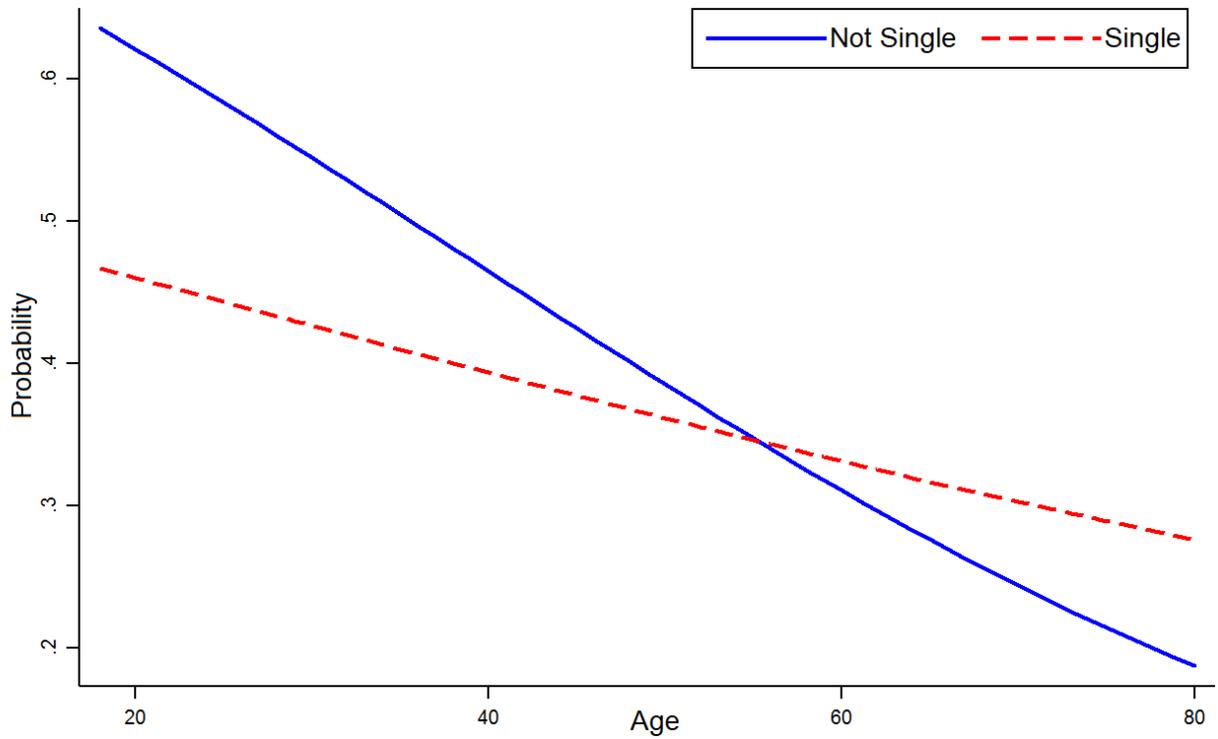

**Figure 16:** Interaction Effect of "Age" and "Single indicator" (Model 4).

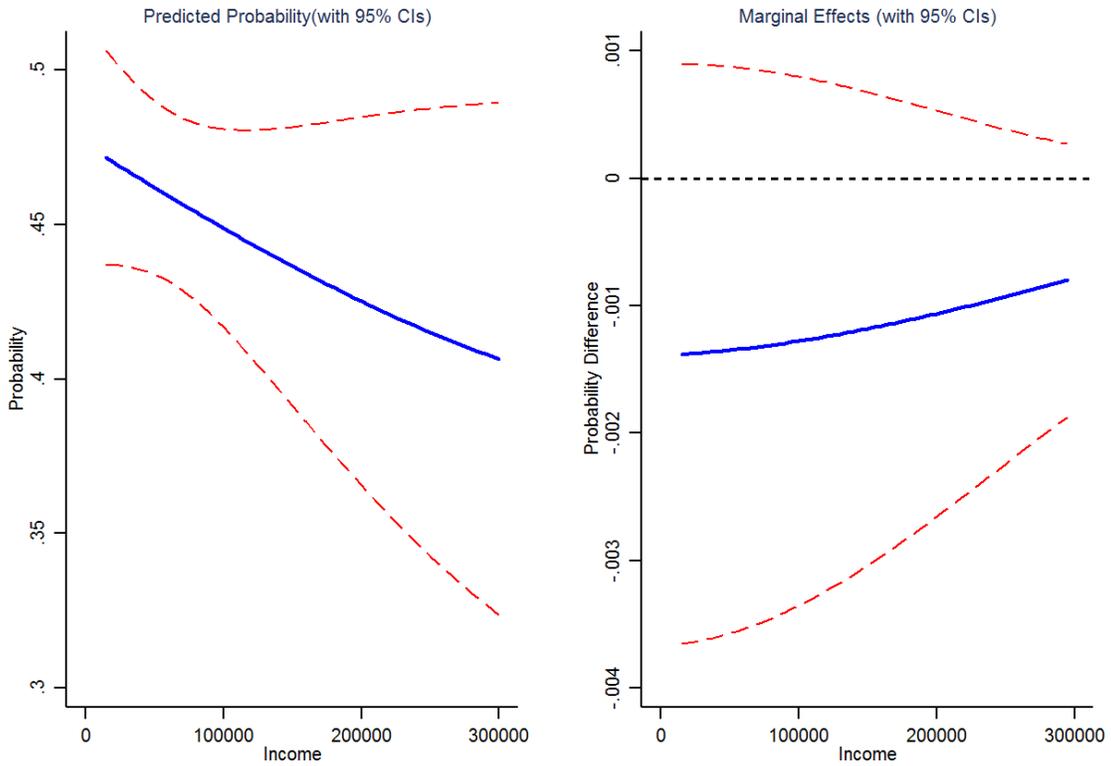

**Figure 17:** Predicted Probability and Marginal Effect of "Annual Income" (Model 4).

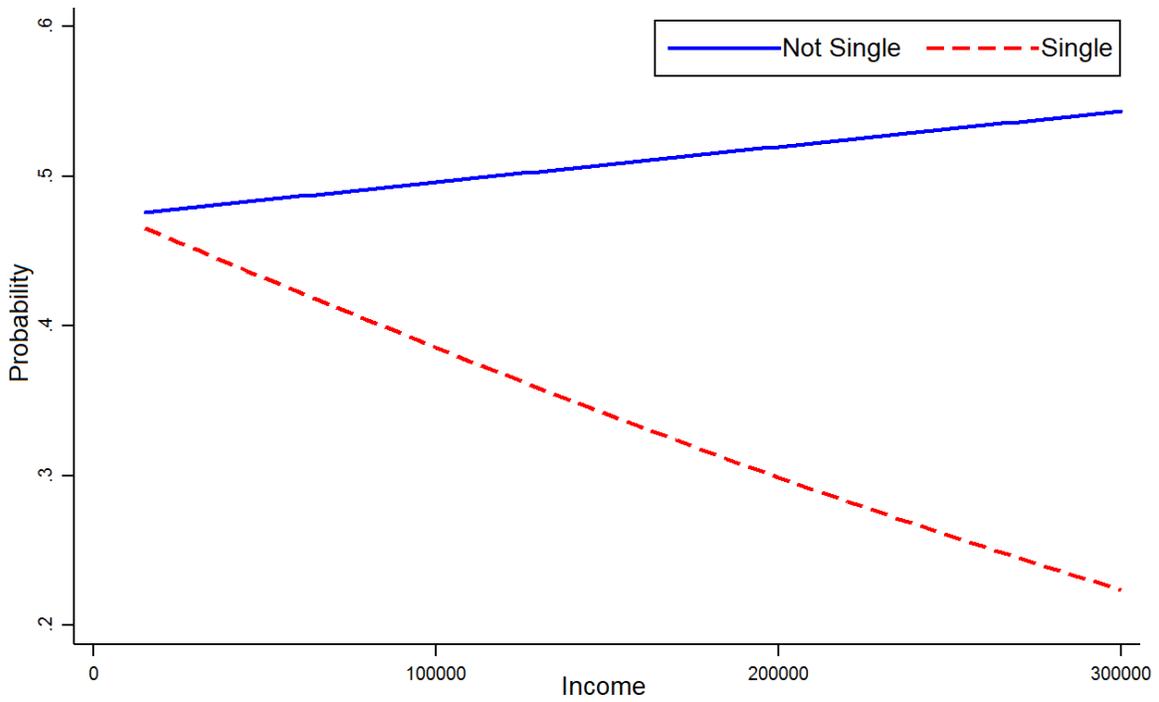

**Figure 18:** Interaction Effect of "Income" and "Single indicator" (Model 4).

# Tables

**Table 1:** Key explanatory variables.

| Explanatory Variables | N | Mean | Median | SD | Min. | Max. |
|---|---|---|---|---|---|---|
| Male indicator | 8791 | 0.48 | 0 | 0.50 | 0 | 1 |
| Single indicator | 8791 | 0.35 | 0 | 0.48 | 0 | 1 |
| Age (in years) | 8791 | 46.77 | 47 | 15.73 | 18 | 100 |
| Postgraduation indicator | 8791 | 0.31 | 0 | 0.46 | 0 | 1 |
| Annual Income (US$) | 8791 | 94909 | 72500 | 90915 | 15000 | 1000000 |
| Metropolitan resident indicator | 8791 | 0.30 | 0 | 0.46 | 0 | 1 |
| Household size 3+ indicator | 8025 | 0.32 | 0 | 0.47 | 0 | 1 |
| Total vehicle ownership | 8791 | 2.15 | 2 | 1.17 | 0 | 6 |
| Early adopter indicator | 8791 | 0.19 | 0 | 0.39 | 0 | 1 |
| Drive daily indicator | 1541 | 0.25 | 0 | 0.43 | 0 | 1 |

**Table 2:** Key response variables.

| Response variables (indicators) | N | Mean | SD |
|---|---|---|---|
| **Model 1** | | | |
| TNC driver | 8791 | 0.29 | 0.45 |
| TNC user | 8791 | 0.29 | 0.45 |
| TNC non-user | 8791 | 0.43 | 0.49 |
| **Model 2** | | | |
| Ridepooling user | 2365 | 0.13 | 0.33 |
| **Model 3** | | | |
| TNC drivers who would prefer to switch to fuel-efficient vehicles | 1533 | 0.53 | 0.50 |
| **Model 4** | | | |
| TNC drivers who considered driving for ridehailing services while buying or leasing a new vehicle | 1540 | 0.47 | 0.50 |

**Table 3:** Most often used mobility options of the ridehailing users (N = 11,902).

| Frequency of using ridehailing services or Taxis | User Type | What mobility options do you use most often? | | | | |
|---|---|---|---|---|---|---|
| | | Ridehailing | Driving personal vehicle | Public Transit | Walk or Bike | Carsharing or carpooling |
| Once or more a day | Frequent TNC Users | 31.88 | 53.49 | 4.26 | 5.62 | 4.74 |
| Once or more per week | | 21.91 | 60.73 | 10.13 | 5.4 | 1.83 |
| Once or more per month; Once or more in 3 months | Infrequent TNC Users | 5.41 | 79.13 | 6.08 | 6.41 | 2.97 |
| Once or more a year; Once or more every few years or Never | | 2.86 | 87.05 | 2.92 | 5.4 | 1.76 |

**Table 4:** Mobility option most used in the absence of their current travel mode.

| Mobility option used most often | N | Ridehailing | Drive personal vehicle | Public transit | Walk or Bike | Carsharing or carpooling | Wouldn't make the trip |
|---|---|---|---|---|---|---|---|
| Ridehailing | 256 | 0 | 65.53 | 14.11 | 6.56 | 13.35 | 0.45 |
| Drive personal vehicle | 6369 | 30.64 | 0 | 9.4 | 7.8 | 46.11 | 6.04 |
| Public transit | 243 | 29.05 | 42.9 | 0 | 16.99 | 3.58 | 1.66 |
| Walk or bike | 372 | 12.42 | 59.84 | 16.88 | 0 | 5.51 | 5.35 |
| Carsharing or carpooling | 151 | 29.85 | 52.21 | 13.34 | 2.17 | 0 | 2.43 |

**Table 5:** Decision to rent/lease/purchase car based on the frequency of driving for ridehailing.

| Frequency of driving for ridehailing services | Number of responses | Was driving for ridehailing service a consideration in the decision to rent/lease/purchase your primary vehicle? | |
|---|---|---|---|
| | | Yes | No |
| Daily | 478 | 65% | 35% |
| Every other day | 343 | 51% | 49% |
| Once in a week | 646 | 43% | 57% |
| Once in a month | 298 | 46% | 54% |
| Less than once per month | 353 | 40% | 60% |

**Table 6:** Multinomial Logistic Parameter Estimates and Relative Risk Ratios (Model 1).

| Explanatory Variables | Parameter estimates | | | Relative Risk ratios | | |
|---|---|---|---|---|---|---|
| | Estimate | Std. Err. | z-stat | Estimate | LB (95% CI) | UB (95% CI) |
| | **Ridehailing Driver** | | | | | |
| Male indicator | -0.530 | 0.100 | -5.28 | 0.589 | 0.484 | 0.717 |
| Single indicator | -0.026 | 0.121 | -0.22 | 0.974 | 0.769 | 1.234 |
| Age | -0.134 | 0.005 | -27.76 | 0.875 | 0.866 | 0.883 |
| Postgraduation indicator | -0.620 | 0.120 | -5.15 | 0.538 | 0.425 | 0.681 |
| Annual Income (US$) | -5.07E-06 | 1.55E-06 | -3.27 | 0.999995 | 0.999992 | 0.999998 |
| Metropolitan resident indicator | 0.665 | 0.112 | 5.93 | 1.945 | 1.561 | 2.423 |
| HH Size 3+ indicator | 0.485 | 0.108 | 4.49 | 1.625 | 1.315 | 2.008 |
| Total vehicle ownership | -0.116 | 0.049 | -2.38 | 0.890 | 0.809 | 0.980 |
| Early adopter indicator | 1.571 | 0.117 | 13.46 | 4.813 | 3.828 | 6.050 |
| Constant | 5.693 | 0.261 | 21.83 | | | |
| | **Ridehailing User** | | | | | |
| Male indicator | -0.286 | 0.072 | -3.97 | 0.751 | 0.652 | 0.865 |
| Single indicator | 0.292 | 0.092 | 3.17 | 1.339 | 1.118 | 1.604 |
| Age | -0.045 | 0.003 | -15.70 | 0.956 | 0.950 | 0.961 |
| Postgraduation indicator | 0.141 | 0.077 | 1.83 | 1.152 | 0.990 | 1.340 |
| Annual Income (US$) | 7.35E-06 | 5.82E-07 | 12.63 | 1.000007 | 1.000006 | 1.000008 |
| Metropolitan resident indicator | 0.425 | 0.088 | 4.83 | 1.529 | 1.287 | 1.817 |
| HH Size 3+ indicator | -0.160 | 0.083 | -1.93 | 0.852 | 0.724 | 1.002 |
| Total vehicle ownership | -0.086 | 0.035 | -2.47 | 0.918 | 0.857 | 0.982 |
| Early adopter indicator | 0.309 | 0.109 | 2.83 | 1.362 | 1.100 | 1.687 |
| Constant | 1.235 | 0.195 | 6.33 | | | |
| | | | | | | |
| N | 8,086 | | | | | |
| Loglikelihood | -6109.5 | | | | | |
| Pseudo R-square | 0.292 | | | | | |

**Note:** "*Ridehailing Non-user*" is a base category. LB (95% CI) and UB (95% CI) imply lower and upper bounds of 95% confidence interval.

**Table 7:** Binary Logistic Parameter Estimates and Odds Ratios (Model 2).

| Explanatory Variables | Parameter estimates | | | Odds ratio | | |
|---|---|---|---|---|---|---|
| | Estimate | Std. Err. | z-stat | Estimate | LB (95% CI) | UB (95% CI) |
| Male indicator | -1.341 | 0.635 | -2.11 | 0.262 | 0.075 | 0.908 |
| Age | -0.047 | 0.014 | -3.43 | 0.955 | 0.929 | 0.980 |
| Postgraduation indicator | -0.614 | 0.662 | -0.93 | 0.541 | 0.148 | 1.981 |
| Metropolitan resident indicator | 0.550 | 0.165 | 3.34 | 1.733 | 1.255 | 2.393 |
| Total vehicle ownership | -0.138 | 0.084 | -1.64 | 0.871 | 0.739 | 1.027 |
| Early adopter indicator | 0.321 | 0.201 | 1.6 | 1.378 | 0.930 | 2.042 |
| Male indicator X Age | 0.026 | 0.014 | 1.89 | 1.026 | 0.999 | 1.054 |
| Postgrad indicator X Age | 0.016 | 0.014 | 1.13 | 1.016 | 0.988 | 1.045 |
| Constant | 0.207 | 0.601 | 0.34 | | | |
| | | | | | | |
| N | 2,504 | | | | | |
| Loglikelihood | -853.9 | | | | | |
| Pseudo R-square | 0.048 | | | | | |

**Note:** "Ride*pooling non-user*" is a base category and parameter estimates are for "Ride*pooling user*". LB (95% CI) and UB (95% CI) imply lower and upper bounds of 95% confidence interval.

**Table 8:** Binary Logistic Parameter Estimates and Odds Ratios (Model 3).

| Explanatory Variables | Parameter estimates | | | Odds ratio | | |
|---|---|---|---|---|---|---|
| | Estimate | Std. Err. | z-stat | Estimate | LB (95% CI) | UB (95% CI) |
| Drive daily indicator | 0.346 | 0.147 | 2.35 | 1.413 | 1.059 | 1.887 |
| Single indicator | -1.026 | 0.431 | -2.38 | 0.359 | 0.154 | 0.834 |
| Age | -0.015 | 0.009 | -1.64 | 0.985 | 0.967 | 1.003 |
| Postgrad indicator | 1.447 | 0.573 | 2.53 | 4.250 | 1.384 | 13.054 |
| Metropolitan resident indicator | 1.392 | 0.441 | 3.16 | 4.024 | 1.695 | 9.554 |
| Total vehicle ownership | 0.054 | 0.055 | 0.97 | 1.055 | 0.947 | 1.176 |
| Early adopter indicator | 0.385 | 0.129 | 2.98 | 1.470 | 1.140 | 1.895 |
| Single indicator X Age | 0.016 | 0.013 | 1.3 | 1.017 | 0.992 | 1.042 |
| Postgrad indicator X Age | -0.032 | 0.016 | -2.02 | 0.969 | 0.940 | 0.999 |
| Metropolitan resident indicator X Age | -0.032 | 0.013 | -2.5 | 0.969 | 0.945 | 0.993 |
| | | | | | | |
| N | 1,534 | | | | | |
| Loglikelihood | -994.3 | | | | | |
| Pseudo R-square | 0.06 | | | | | |

**Note:** "*No preference of ridehailing drivers to switch to fuel efficient vehicles*" is a base category and parameters are estimated for "P*reference of ridehailing…*". LB (95% CI) and UB (95% CI) imply lower and upper bounds of 95% confidence interval.

**Table 9:** Binary Logistic Parameter Estimates and Odds Ratios (Model 4).

| Explanatory Variables | Parameter estimates | | | Odds ratio | | |
|---|---|---|---|---|---|---|
| | **Estimate** | **Std. Err.** | **z-stat** | **Estimate** | **LB (95% CI)** | **UB (95% CI)** |
| Drive daily indicator | 0.736 | 0.159 | 4.620 | 2.09 | 1.53 | 2.85 |
| Male indicator | 0.459 | 0.120 | 3.840 | 1.58 | 1.25 | 2.00 |
| Single indicator | -0.518 | 0.529 | -0.980 | 0.60 | 0.21 | 1.68 |
| Age | -0.032 | 0.010 | -3.150 | 0.97 | 0.95 | 0.99 |
| Postgraduation indicator | 1.122 | 0.588 | 1.910 | 3.07 | 0.97 | 9.72 |
| Annual Income (US$) | 1.17E-06 | 6.18E-07 | 1.900 | 1.000001 | 1 | 1.000002 |
| Metropolitan resident indicator | 0.639 | 0.138 | 4.620 | 1.90 | 1.44 | 2.49 |
| Total vehicle ownership | 0.163 | 0.056 | 2.920 | 1.18 | 1.05 | 1.31 |
| Early adopter indicator | 0.953 | 0.135 | 7.060 | 2.59 | 1.99 | 3.38 |
| Single indicator X Age | 0.020 | 0.016 | 1.250 | 1.02 | 0.99 | 1.05 |
| Postgrad indicator X Age | -0.024 | 0.017 | -1.400 | 0.98 | 0.95 | 1.01 |
| Single indicator X Annual Income | -5.64E-06 | 2.42E-06 | -2.330 | 0.999994 | 0.999990 | 0.999999 |
| Constant | -0.699 | 0.376 | -1.860 | | | |
| | | | | | | |
| N | 1,539 | | | | | |
| Loglikelihood | -914.5 | | | | | |
| Pseudo R-square | 0.139 | | | | | |

**Note:** "*No change* in p*reference to buy/lease/rent a new vehicle with driving for TNCs being a major consideration*" is a base category and parameters are estimated for "*change in preference…*". LB (95% CI) and UB (95% CI) imply lower and upper bounds of 95% confidence interval.